\documentstyle[12pt]{article}

\begin{document}

\begin{center}
{\Large{\bf Fixed Points in Self--Similar Analysis of Time Series} \\ [5mm]
S. Gluzman$^1$ and V.I. Yukalov$^2$} \\ [3mm]
{\it $^1$International Centre of Condensed Matter Physics \\
University of Brasilia, Brasilia, DF 70919--970, Brazil \\
and \\
$^2$ Centre for Interdisciplinary Studies in Chemical Physics \\
University of Western Ontario, London, Ontario N6A 3K7, Canada}
\end{center}

\vspace{2cm}

\begin{abstract}

Two possible definitions of fixed points in the self--similar analysis of
time series are considered. One definition is based on the
minimal--difference condition and another, on a simple averaging. From
studying stock market time series, one may conclude that these two
definitions are practically equivalent. A forecast is made for the stock
market indices for the end of March 1998.

\end{abstract}

\vspace{2cm}

Time series analysis and forecasting have a long history and abundant
literature, to mention just a few Refs. [1-3]. When analysing time series,
one usually aims at constructing a particular model that could represent
the available historical data and, after such a model is constructed, one
could use it for predicting future. This kind of approach has been found
rather reasonable for describing sufficiently smooth evolution, but it
fails in treating large fluctuations, like those happenning in stock
markets. This failure is caused by quite irregular evolution of markets
whose calm development is often interrupted by sudden strong deviations
producing booms and crushes. Such deviations are not regular cyclic
oscillations but rather are chaotic events, alike heterophase fluctuations
in statistical systems [4]. Similarly to the latter, strong market
fluctuations are also of coherent nature, having their origin in the
collective nonlinear interactions of many trading agents. The coherent
collective behaviour of traders is often termed the crowd or herd
behaviour [5-7]. Strong nonlinearity and nonequilibrium of stock markets
make them one of the most complex systems existing in nature, comparable
with human brain.

A novel approach to analysing and forecasting time series has been
recently suggested [8,9]. Being based on the self--similar approximation
theory [10-17], this technique can be called the self--similar analysis
of time series. In this approach, instead of trying to construct a
particular model imitating the dynamical system generating time series, we
assume that the evolution of the system is self--similar. This is almost
the same as to say that the dynamics of the considered system is governed
by some laws. Since the observed time series data are the product of a
self--similar evolution, the information on some kind of self--similarity
is to be hidden in these data. The role of the self--similar analysis is
to extract this hidden information.

We applied the self--similar analysis to stock market time series in Refs.
[8,9], where we used two definitions of fixed points resulting in two
possible forecasts, $f_n^*(n+1)$ and $\stackrel{-}{f}_n(n+1)$. Here and in
what follows we use the notation of Ref. [9]. The aim of the present
letter is to pay a special attention to comparing these two ways of
defining fixed points. We consider stock market indices for the cases when
the answer is known and also make predictions for the end of March 1998.

\vspace{3mm}

{\bf NYSE Composite}. Let us try to make a forecast for the end of
{\sl February 1998}. The following data are available in the period of
time from {\sl August 31, 1997} till {\sl January 31, 1998}, taken with
one
month resolution: 
$$
470.48\; (Aug.\; 31),\quad 497.23,\quad 481.14,\quad 499.10,\quad
511.19,\quad 510.63\ (Jan.\; 31). 
$$
The self--similar exponential approximations and corresponding multipliers
can be obtained: 
$$
f_2^{*}(3)=531.074,\quad f_3^{*}(4)=563.269,\quad f_4^{*}(5)=449.993,\quad
f_5^{*}(6)=550.72 , 
$$
$$
\stackrel{-}{f}_2(3)=538.364,\quad \stackrel{-}{f}_3(4)=566.264,\quad
\stackrel{-}{f}_4(5)=438.115,\quad \stackrel{-}{f}_5(6)=539.475 ,
$$
$$
M_2^{*}(3)=0.38,\quad M_3^{*}(4)=1.022,\quad M_4^{*}(5)=0.215,\quad 
\left| M_5^{*}(6)\right| =0.013, 
$$
$$
\stackrel{-}{M}_2(3)=0.495,\quad \stackrel{-}{M}_3(4)=1.115,\quad
\stackrel{-}{M}_4(5)=0.347,\quad 
\left | \stackrel{-}{M}_5(6)\right | =0.005. 
$$
By the end of February the index was $544.26$, which should be compared
with $\stackrel{-}{f}_5(6)$. Let us make a forecast for the end of
{\sl March, 1998}:
$$
f_2^{*}(3)=492.69,\quad f_3^{*}(4)=518.884,\quad f_4^{*}(5)=8042,\quad
f_5^{*}(6)=425.835 , 
$$
$$
\stackrel{-}{f}_2(3)=484.587,\quad \stackrel{-}{f}_3(4)=522.012,\quad
\stackrel{-}{f}_4(5)=2.127\times 10^{14},\quad 
\stackrel{-}{f}_5(6)=420.252 ,
$$
$$
M_2^{*}(3)=0.184,\quad \left | M_3^{*}(4)\right | =0.042,\quad 
\left| M_4^{*}(5)\right | =385,\quad M_5^{*}(6)=0.605 ,
$$
$$
\stackrel{-}{M}_2(3)=0.492,\quad \stackrel{-}{M}_3(4)=0.1 ,\quad
\stackrel{-}{M}_4(5)=1.12\times 10^{15},\quad 
\stackrel{-}{M}_5(6)=0.726. 
$$
The optimal forecast is $f_3^*(4)$.

{\bf S\&P}. Let us try to make a forecast for the end of {\sl February 1998}.
The following data are available in the period of time from {\sl August
31, 1997}  till {\sl January 31, 1998}, taken with one month resolution: 
$$
900\; ({\sl Aug.\; 31}), \qquad 950,\qquad 915, \qquad 955,\qquad 970,
\qquad 980\; ({\sl Jan.\;31}). 
$$
The self--similar exponential approximations and corresponding multipliers
can be obtained: 
$$
f_2^{*}(3)=994.161,\quad f_3^{*}(4)=1004,\quad f_4^{*}(5)=832.896,\quad
f_5^{*}(6)=1058 ,
$$
$$
\stackrel{-}{f}_2(3)=999.82,\quad \stackrel{-}{f}_3(4)=995.585,\quad
\stackrel{-}{f}_4(5)=798.7576,\quad \stackrel{-}{f}_5(6)=1038 ,
$$
$$
M_2^{*}(3)=0.607,\quad M_3^{*}(4)=0.055,\quad M_4^{*}(5)=0.286,\quad
\left| M_5^{*}(6)\right| =0.0115 ,
$$
$$
\stackrel{-}{M}_2(3)=0.683,\quad \stackrel{-}{M}_3(4)=0.076,\quad
\stackrel{-}{M}_4(5)=0.455,\quad \left | \stackrel{-}{M}_5(6)\right | =0.0025. 
$$
By the end of February the index was $1049$, which should be compared with 
$\stackrel{-}{f}_5(6)$. Let us make a forecast for the end of {\sl March,
1998}:
$$
f_2^{*}(3)=953.742,\quad f_3^{*}(4)=979.487,\quad f_4^{*}(5)=915,\quad
f_5^{*}(6)=862.455 ,
$$
$$
\stackrel{-}{f}_2(3)=941.302,\quad \stackrel{-}{f}_3(4)=985.725,\quad
\stackrel{-}{f}_4(5)=915,\quad \stackrel{-}{f}_5(6)=874.712 ,
$$
$$
M_2^{*}(3)=0.106,\quad \left | M_3^{*}(4)\right | =0.035,\quad
M_4^{*}(5)\approx 0,\quad \left | M_5^{*}(6)\right | =0.104 ,
$$
$$
\stackrel{-}{M}_2(3)=0.48,\quad \stackrel{-}{M}_3(4)=0.169,\quad
\stackrel{-}{M}_4(5)\approx 0,\quad
\left | \stackrel{-}{M}_5(6)\right | =0.105 .
$$
The optimal forecasts are $f_4^*(5)=\stackrel{-}{f}_4(5)$.

{\bf Dow Jones}. Let us try to make a forecast for the end of
{\sl February 1998}. The following data are available in the period of
time from {\sl August 31, 1997} till {\sl January 31, 1998}, taken with
one month resolution: 
$$
7670\; ({\sl Aug.\; 31}), \qquad 8000,\qquad 7442, \qquad 7823,\qquad 7908,
\qquad 7907\; ({\sl Jan.\; 31}) . 
$$
The self--similar exponential approximations and corresponding multipliers
can be obtained: 
$$
f_2^{*}(3)=8043,\quad f_3^{*}(4)=8235,\quad f_4^{*}(5)=6347,\quad
f_5^{*}(6)=9069 ,
$$
$$
\stackrel{-}{f}_2(3)=8091,\quad \stackrel{-}{f}_3(4)=8152,\quad
\stackrel{-}{f}_4(5)=5965, \quad \stackrel{-}{f}_5(6)=8898 ,  
$$
$$
M_2^{*}(3)=0.396,\quad M_3^{*}(4)=0.015,\quad M_4^{*}(5)=0.436,\quad
\left | M_5^{*}(6)\right | =0.007 , 
$$
$$
\stackrel{-}{M}_2(3)=0.499,\quad \stackrel{-}{M}_3(4)=0.038,\quad
\stackrel{-}{M}_4(5)=0.605,\quad \left | 
\stackrel{-}{M}_5(6)\right | =0.0015 .
$$
By the end of February the index was $8546$, which should be compared with 
$\stackrel{-}{f}_5(6)$. Let us make a forecast for the end of {\sl March,
1998}:
$$
f_2^{*}(3)=7578,\quad f_3^{*}(4)=7968,\quad f_4^{*}(5)=7442,\quad
f_5^{*}(6)=6721 ,
$$
$$
\stackrel{-}{f}_2(3)=7431,\quad \stackrel{-}{f}_3(4)=8032,\quad
\stackrel{-}{f}_4(5)=7442,\quad \stackrel{-}{f}_5(6)=6890 ,
$$
$$
M_2^*(3)=0.18,\quad \left | M_3^*(4)\right | =0.049,
\quad M_4^*(5)\approx 0,\quad \left | M_5^*(6)\right | =0.15 ,
$$
$$
\stackrel{-}{M}_2(3)=0.444,\quad \stackrel{-}{M}_3(4)=0.127,\quad
\stackrel{-}{M}_4(5)\approx 0,\quad \left | 
\stackrel{-}{M}_5(6)\right | =0.168 .
$$
The optimal forecasts are $f_4^*(5)=\stackrel{-}{f}_4(5)$.

{\bf Nasdaq Composite}. Let us try to make a forecast for the end of
{\sl February 1998}. The following data are available in the period of
time from {\sl August 31, 1997} till {\sl January 31, 1998}, taken with
one month resolution: 
$$
1600\; ({\sl Aug.\; 31}), \qquad 1690,\qquad 1594,\qquad 1601,\qquad 1570,
\qquad 1619\; ({\sl Jan.\; 31}). 
$$
The self--similar exponential approximations and corresponding multipliers
can be obtained: 
$$
f_2^{*}(3)=1506,\quad f_3^{*}(4)=1612,\quad f_4^{*}(5)=1391,\quad
f_5^{*}(6)=1878 ,
$$
$$
\stackrel{-}{f}_2(3)=1482,\quad \stackrel{-}{f}_3(4)=1618,\quad
\stackrel{-}{f}_4(5)=1323,\quad \stackrel{-}{f}_5(6)=1843 ,
$$
$$
M_2^{*}(3)=0.302,\quad \left| M_3^{*}(4)\right | =0.0535,\quad
M_4^{*}(5)=0.433,\quad \left| M_5^{*}(6)\right | =0.011 ,
$$
$$
\stackrel{-}{M}_2(3)=0.429,\quad \stackrel{-}{M}_3(4)=0.064,\quad
\stackrel{-}{M}_4(5)=0.567,\quad \left | 
\stackrel{-}{M}_5(6)\right | =0.0025. 
$$
By the end of February the index was $1771$, which should be compared with
$\stackrel{-}{f}_5(6)$. Let us make a forecast for the end of {\sl March,
1998}:
$$
f_2^{*}(3)=1569.6,\quad f_3^{*}(4)= 1560,\quad f_4^{*}(5)=1663,\quad
f_5^{*}(6)=1477 ,
$$
$$
\stackrel{-}{f}_2(3)=1566,\quad \stackrel{-}{f}_3(4)=1575,\quad
\stackrel{-}{f}_4(5)=1693,\quad \stackrel{-}{f}_5(6)=1509 ,
$$
$$
\left | M_2^{*}(3)\right | =0.025,\quad \quad M_3^{*}(4)= -0.151 , \quad
M_4^{*}(5)=0.137,\quad \left | M_5^{*}(6)\right | =0.188 , 
$$
$$
\stackrel{-}{M}_2(3)=0.5,\quad \left | \stackrel{-}{M}_3(4)\right | =0.175,
\quad \stackrel{-}{M}_4(5)=0.257,\quad \left | 
\stackrel{-}{M}_5(6)\right | =0.105. 
$$
The optimal forecast is $f_2^*(3)$.

As follows from the analysis of forecasts for February, two ways of
defining fixed points and leading to $f_n^*(n+1)$ or
$\stackrel{-}{f}_n(n+1)$, respectively, are practically equivalent, the
corresponding optimal forecasts being close to each other. How the
forecast for March works, we shall check in a month.

\end{document}